\documentclass[twocolumn]{aastex63}
\usepackage{soul}
\usepackage{soul}
\soulregister\cite7 % 针对\cite命令
\soulregister\citep7 % 针对\citep命令
\soulregister\citet7 % 针对\citet命令
\soulregister\ref7 % 针对\ref命令
\soulregister\pageref7 % 针对\pageref命令

\begin{document}

%%%%%%%%%%%%%%%%%%%%%%%%%%%%%%%Title%%%%%%%%%%%%%%%%%%%%%%%%%%%%%%%%%%%%%
\title{The Background Model of the Medium Energy X-ray telescope of \textit{Insight-HXMT}}
%\author{test \altaffilmark{1},
%Yu-Peng Chen\altaffilmark{1},
%Shu Zhang\altaffilmark{1},
%Shuang-Nan Zhang\altaffilmark{1},\\
%Long Ji\altaffilmark{1},
%Jian Li\altaffilmark{1},
%}
%\correspondingauthor{ChengCheng GUO, JinYuan LIAO}
%\email{guocc@ihep.ac.cn, liaojinyuan@ihep.ac.cn}

\author{ChengCheng GUO}
\altaffiliation{Corresponding author
\\Email addresses guocc@ihep.ac.cn, liaojinyuan@ihep.ac.cn}
\affiliation{Key Laboratory for Particle Astrophysics, Institute of High Energy Physics, Chinese Academy of Sciences, 19B Yuquan Road, Beijing 100049, China}
\affiliation{University of Chinese Academy of Sciences, Chinese Academy of Sciences, Beijing 100049, China}

\author{JinYuan LIAO}
\altaffiliation{Corresponding author
\\Email addresses guocc@ihep.ac.cn, liaojinyuan@ihep.ac.cn}
\affiliation{Key Laboratory for Particle Astrophysics, Institute of High Energy Physics, Chinese Academy of Sciences, 19B Yuquan Road, Beijing 100049, China}

\author{Shu ZHANG}
\affiliation{Key Laboratory for Particle Astrophysics, Institute of High Energy Physics, Chinese Academy of Sciences, 19B Yuquan Road, Beijing 100049, China}

\author{Juan ZHANG}
\affiliation{Key Laboratory for Particle Astrophysics, Institute of High Energy Physics, Chinese Academy of Sciences, 19B Yuquan Road, Beijing 100049, China}

\author{Ying TAN}
\affiliation{Key Laboratory for Particle Astrophysics, Institute of High Energy Physics, Chinese Academy of Sciences, 19B Yuquan Road, Beijing 100049, China}

\author{LiMing SONG}
\affiliation{Key Laboratory for Particle Astrophysics, Institute of High Energy Physics, Chinese Academy of Sciences, 19B Yuquan Road, Beijing 100049, China}

\author{FangJun LU}
\affiliation{Key Laboratory for Particle Astrophysics, Institute of High Energy Physics, Chinese Academy of Sciences, 19B Yuquan Road, Beijing 100049, China}

\author{XueLei CAO}
\affiliation{Key Laboratory for Particle Astrophysics, Institute of High Energy Physics, Chinese Academy of Sciences, 19B Yuquan Road, Beijing 100049, China}

\author{Zhi CHANG}
\affiliation{Key Laboratory for Particle Astrophysics, Institute of High Energy Physics, Chinese Academy of Sciences, 19B Yuquan Road, Beijing 100049, China}

\author{YuPeng CHEN}
\affiliation{Key Laboratory for Particle Astrophysics, Institute of High Energy Physics, Chinese Academy of Sciences, 19B Yuquan Road, Beijing 100049, China}
\author{YuanYuan DU}
\affiliation{Key Laboratory for Particle Astrophysics, Institute of High Energy Physics, Chinese Academy of Sciences, 19B Yuquan Road, Beijing 100049, China}
\author{MingYu GE}
\affiliation{Key Laboratory for Particle Astrophysics, Institute of High Energy Physics, Chinese Academy of Sciences, 19B Yuquan Road, Beijing 100049, China}
\author{YuDong GU}
\affiliation{Key Laboratory for Particle Astrophysics, Institute of High Energy Physics, Chinese Academy of Sciences, 19B Yuquan Road, Beijing 100049, China}
\author{WeiChun JIANG}
\affiliation{Key Laboratory for Particle Astrophysics, Institute of High Energy Physics, Chinese Academy of Sciences, 19B Yuquan Road, Beijing 100049, China}
\author{Jing JIN}
\affiliation{Key Laboratory for Particle Astrophysics, Institute of High Energy Physics, Chinese Academy of Sciences, 19B Yuquan Road, Beijing 100049, China}
\author{Gang LI}
\affiliation{Key Laboratory for Particle Astrophysics, Institute of High Energy Physics, Chinese Academy of Sciences, 19B Yuquan Road, Beijing 100049, China}
\author{Xian LI}
\affiliation{Key Laboratory for Particle Astrophysics, Institute of High Energy Physics, Chinese Academy of Sciences, 19B Yuquan Road, Beijing 100049, China}
\author{XiaoBo LI}
\affiliation{Key Laboratory for Particle Astrophysics, Institute of High Energy Physics, Chinese Academy of Sciences, 19B Yuquan Road, Beijing 100049, China}
\author{ShaoZhen LIU}
\affiliation{Key Laboratory for Particle Astrophysics, Institute of High Energy Physics, Chinese Academy of Sciences, 19B Yuquan Road, Beijing 100049, China}
\author{XiaoJing LIU}
\affiliation{Key Laboratory for Particle Astrophysics, Institute of High Energy Physics, Chinese Academy of Sciences, 19B Yuquan Road, Beijing 100049, China}
\author{XueFeng LU}
\affiliation{Key Laboratory for Particle Astrophysics, Institute of High Energy Physics, Chinese Academy of Sciences, 19B Yuquan Road, Beijing 100049, China}
\author{Tao LUO}
\affiliation{Key Laboratory for Particle Astrophysics, Institute of High Energy Physics, Chinese Academy of Sciences, 19B Yuquan Road, Beijing 100049, China}
\author{Bin MENG}
\affiliation{Key Laboratory for Particle Astrophysics, Institute of High Energy Physics, Chinese Academy of Sciences, 19B Yuquan Road, Beijing 100049, China}
\author{Liang SUN}
\affiliation{Key Laboratory for Particle Astrophysics, Institute of High Energy Physics, Chinese Academy of Sciences, 19B Yuquan Road, Beijing 100049, China}
\author{JiaWei YANG}
\affiliation{Key Laboratory for Particle Astrophysics, Institute of High Energy Physics, Chinese Academy of Sciences, 19B Yuquan Road, Beijing 100049, China}
\author{Sheng YANG}
\affiliation{Key Laboratory for Particle Astrophysics, Institute of High Energy Physics, Chinese Academy of Sciences, 19B Yuquan Road, Beijing 100049, China}
\author{Yuan YOU}
\affiliation{Key Laboratory for Particle Astrophysics, Institute of High Energy Physics, Chinese Academy of Sciences, 19B Yuquan Road, Beijing 100049, China}
\affiliation{University of Chinese Academy of Sciences, Chinese Academy of Sciences, Beijing 100049, China}
\author{WanChang ZHANG}
\affiliation{Key Laboratory for Particle Astrophysics, Institute of High Energy Physics, Chinese Academy of Sciences, 19B Yuquan Road, Beijing 100049, China}
\author{HaiSheng ZHAO}
\affiliation{Key Laboratory for Particle Astrophysics, Institute of High Energy Physics, Chinese Academy of Sciences, 19B Yuquan Road, Beijing 100049, China}
\author{ShuangNan ZHANG}
\affiliation{Key Laboratory for Particle Astrophysics, Institute of High Energy Physics, Chinese Academy of Sciences, 19B Yuquan Road, Beijing 100049, China}
\affiliation{University of Chinese Academy of Sciences, Chinese Academy of Sciences, Beijing 100049, China}
\affiliation{Key Laboratory of Space Astronomy and Technology, National Astronomical Observatories, Chinese Academy of Sciences, Beijing 100012, China}

%\author{Cheng-Cheng Guo\altaffilmark{1,2,*}, Jin-Yuan Liao\altaffilmark{1,*}, Shu Zhang\altaffilmark{1}, Shuang-Nan Zhang\altaffilmark{1,2,3}, et al. \\ \emph{Insight-HXMT} Group}

%\affil{$^{1}$Key Laboratory for Particle Astrophysics, Institute of High Energy Physics, Chinese Academy of Sciences, 19B Yuquan Road, Beijing 100049, China}
%\altaffiltext{1}{Key Laboratory for Particle Astrophysics, Institute of High Energy Physics, Chinese Academy of Sciences, 19B Yuquan Road, Beijing 100049, China}
%\altaffiltext{2}{University of Chinese Academy of Sciences, Chinese Academy of Sciences, Beijing 100049, China}
%\altaffiltext{3}{Key Laboratory of Space Astronomy and Technology, National Astronomical Observatories, Chinese Academy of Sciences, Beijing 100012, China}

%\altaffiltext{1}
%{Key Laboratory for Particle Astrophysics, Institute of High Energy Physics, Chinese Academy of Sciences, 19B Yuquan Road, Beijing 100049, China}

%%%%%%%%%%%%%%%%%%%%%%%%%%%%%%% Abstract %%%%%%%%%%%%%%%%%%%%%%%%%%%%%%%%%
\begin{abstract}
The Medium Energy X-ray Telescope (ME) is one of the main payloads of the \emph{Hard X-ray Modulation Telescope} (dubbed as \emph{Insight-HXMT}). The background of \emph{Insight-HXMT}/ME is mainly caused by the environmental charged particles and the background intensity is modulated remarkably by the geomagnetic field, as well as the geographical location. At the same geographical location, the background spectral shape is stable but the intensity varies with the level of the environmental charged particles. In this paper, we develop a model to estimate the ME background based on the ME database that is established with the two-year blank sky observations of the high Galactic latitude. In this model, the entire geographical area covered by \emph{Insight-HXMT} is divided into grids of $5^{\circ}\times5^{\circ}$ in geographical coordinate system. For each grid, the background spectral shape can be obtained from the background database and the intensity can be corrected by the contemporary count rate of the blind FOV detectors. Thus the background spectrum can be obtained by accumulating the background of all the grids passed by \emph{Insight-HXMT} during the effective observational time. The model test with the blank sky observations shows that the systematic error of the background estimation in $8.9-44.0$ keV is $\sim1.3\%$ for a pointing observation with an average exposure $\sim5.5$ ks. We also find that the systematic error is anti-correlated with the exposure, which indicates the systematic error is partly contributed by the statistical error of count rate measured by the blind FOV detectors.
\end{abstract}

%%%%%%%%%%%%%%%%%%%%%%%%%%%%%%%%% Section 1 %%%%%%%%%%%%%%%%%%%%%%%%%%%%%%%
\section{Introduction}

The Medium-Energy X-ray Telescope (ME, Cao et al. 2020) is one of the main payloads of the \emph{Hard X-ray Modulation Telescope} (dubbed as \emph{Insight-HXMT}), 
which is China's first X-ray astronomical satellite launched on June 15, 2017 (Zhang et al. 2020). The task of ME module is to provide measurements of X-ray sources
in $5-40$~keV and, together with the measurements with the Low Energy X-ray Telescope (LE, Chen et al. 2020) and the High Energy X-ray Telescope (HE, Liu et al. 2020), to study these sources in a rather broad energy band of $1-250$ keV. 
It contains three detector boxes and each detector box has three FPGA (Field Programmable Gate Array) modules. 
As described in Cao et al. (2020), each FPGA operates six ASIC (Application Specific Integrated Circuit) modules, each of which handles 32 SI-PIN pixels. 
In total, ME consists of 1728 Si-PIN pixels with an energy range of $5-40$ keV and a total geometrical area of $952~{\rm cm}^2$.
%ME takes the same strategy on FOV as that of LE and HE.
The orientations of the field of views (FOVs) of the three detector boxes differ by 60$^\circ$.
Each detector box has 15 ASICs for small FOV (1$^\circ\times4^\circ$) pixels and two ASICs for large FOV (4$^\circ\times4^\circ$) pixels.
In addition, there is also one ASIC for blind FOV pixels, i.e., the collimator is blocked on top by an aluminium plate.
Hereafter the ``small FOV detector" refers to all the pixels of the 45 small FOV ASICs, the ``large FOV detector" refers to all the pixels of the six large FOV ASICs and the ``blind FOV detector" refers to all the pixels of the three blind ASICs for convenience.

%An accurate background estimation is necessary for both the temporal and spectral analysis. 
The typical flux of ME background is $\sim130$~mCrab and it varies with geographical location by a factor of 3. 
For observation of a weak X-ray source $\sim10$~mCrab with an exposure longer than 1~ks, 
in order to have spectral or timing result with significance $>5\sigma$, the background estimation should be as accurate as better than $3\%$. 
As a collimated telescope, ME does not have direct imaging capability or works in a rocking mode for direct background measurement. 
It is thus essential to develop a background model based on the characteristics of ME background observations in order to obtain an accurate background estimation.

%new1%
Because \emph{Insight-HXMT}/ME has a unique design of the blind detector, a simple way is to use the blind detector as a background detector, and use the blank sky observation to obtain correlation between the small FOV detector and the blind detector.
However, the EC relationship and the energy resolution are both different between the blind and small/large FOV detectors, which result in the obvious residual structure in the background estimation. In addition, the number of blind detectors is too small (only 1/18 of total ME pixels), thus the statistical error is also very large.

\begin{figure}
    \centering
    \includegraphics[scale=0.4]{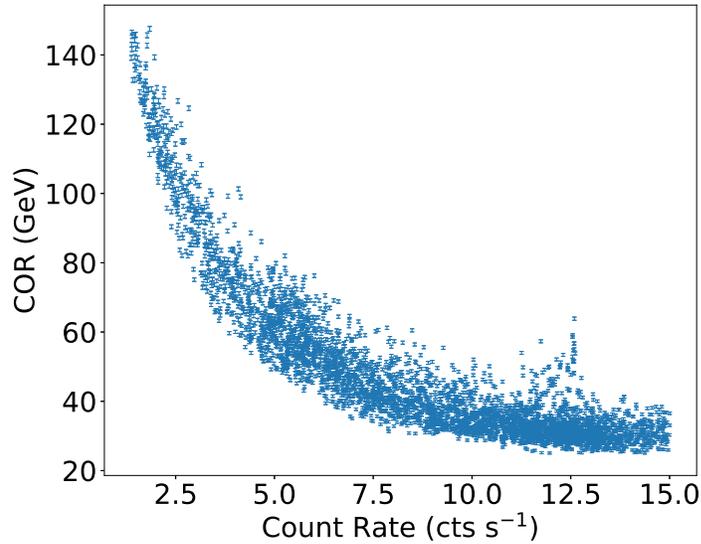}
    \caption{The relationship between COR and the background count rates of small FOV detectors in $8.9-44.0$ keV. Each point has an exposure $300$~s.}
    \label{cor_small_relation}
\end{figure}

Since the background intensity is modulated remarkably by the geomagnetic field and the geographical location (Section 3), the parameters ralated to the geomagnetic field can be used to estimate the ME background. We investigate the correlation between the geomagnetic cut-off rigidity (COR), which is inversely proportional to the square of the McIlwain Parameter $L$ in McIlwain coordinate systems (Dean et al. 2003), and ME backgrounds. As shown in Figure~\ref{cor_small_relation}, the large dispersion means that there is large uncertainty in the background estimation by taking only COR or $L$.

%new1%
In the adopted model, the entire geographical area covered by \emph{Insight-HXMT} is divided into hundreds of small grids
($5^{\circ}\times5^{\circ}$ in geographical coordinate system). The spectral shape of the background is stable for each grid, thus for a pointing observation the shape of the background spectrum can be determined by the geographical regions passed by the satellite.
In addition, the intensity of the background spectrum can be corrected with contemporary measurement of the blind FOV detector.

In this work, we first study the main characteristics of the ME background, and then develop a ME background model with the blank sky observations.
Finally, the model test is performed with the blank sky observations that are not relevant to the model construction. Currently, the ME background model described here has been adopted as the standard method to estimate the ME background in \emph{Insight-HXMT} data analysis software HXMTDAS.
This paper is organized as follows. The characteristics of the ME background are presented in Section 2. 
The principle of the background model is described in Section 3. The result of the background reproducibility is presented in Section 4. 
Finally, the discussion and summary are given in Sections 5 and 6, respectively.

\section{Data Reduction}
The data reduction of \emph{Insight-HXMT}/ME is composed of three parts: the first is the preliminary data reduction 
by the \emph{Insight-HXMT} data analysis software package HXMTDAS v2.0, the second is further data reduction to remove 
the special abnormal time from the good time interval (GTI), and the third is the good pixel dynamic selection.

The preliminary data reduction is performed by the \emph{Insight-HXMT} data analysis software package HXMTDAS v2.0 (HXMT User Analysis Software Group 2019) and goes through the following steps: PI (pulse invariant) transformation, grade calculation, and basic GTI selection. 
The PI transformation is performed first and then the event grade is calculated. In this paper, we only use the events with grade 0, which means the coincidence events are removed because they are usually caused by charged particles rather than X-ray photons. For the GTI selection, there are a variety of criteria used to obtain clean data. The details of these criteria are shown in Table 1.

\begin{figure}
    \centering
    \includegraphics[scale=0.4]{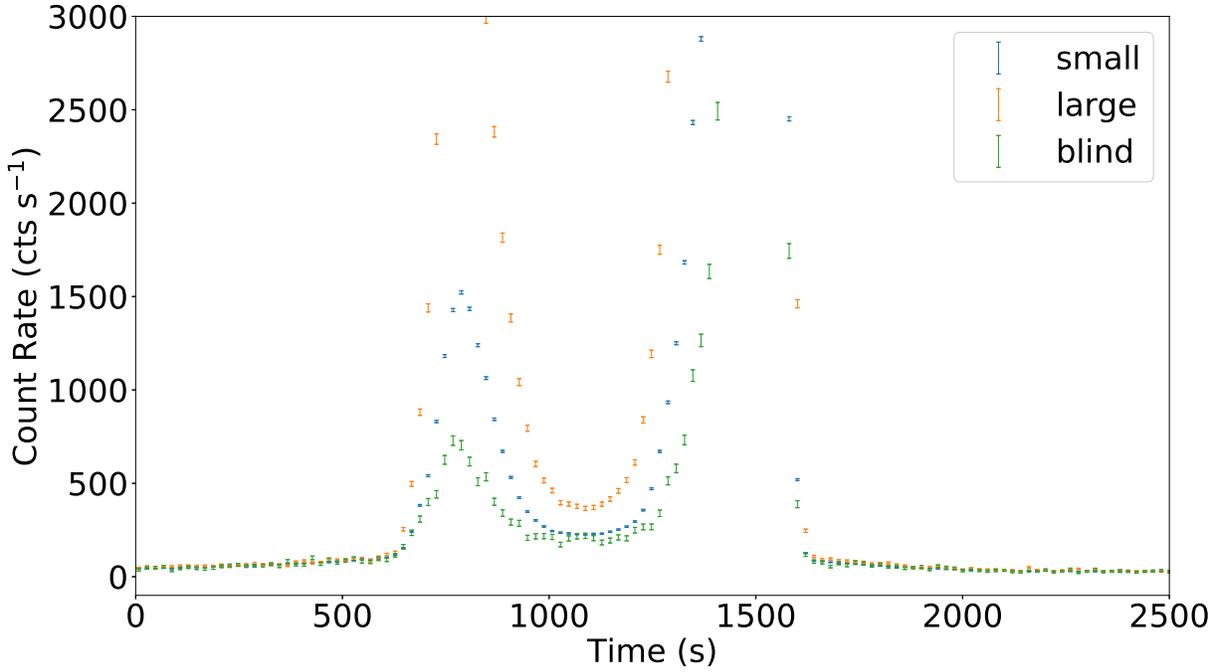}
    \caption{An example of a \emph{Insight-HXMT}/ME background light curve with $\rm T_{bin}=20$~s of the detectors with the blind, small and large FOVs in $8.9-44.0$ keV. The fluxes of the large and blind FOV detectors are normalized to that as the same pixel numbers as the small FOV detector. The intensities of the peak flares increase with the FOVs of the detectors.}
    %This figure shows a very strong flare but sometimes background flare is much weaker than normal background and hard to recognize.}
    \label{figgrosslc}
\end{figure}

\begin{figure}
    \centering
    \includegraphics[scale=0.4]{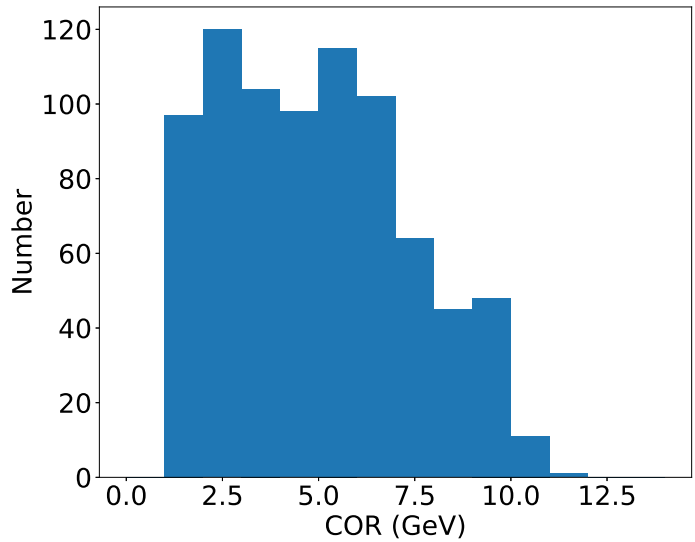}
    \caption{COR distribution of the background flares.}
    \label{flare_distribution}
\end{figure}

\begin{figure}
    \centering
    \includegraphics[scale=0.4]{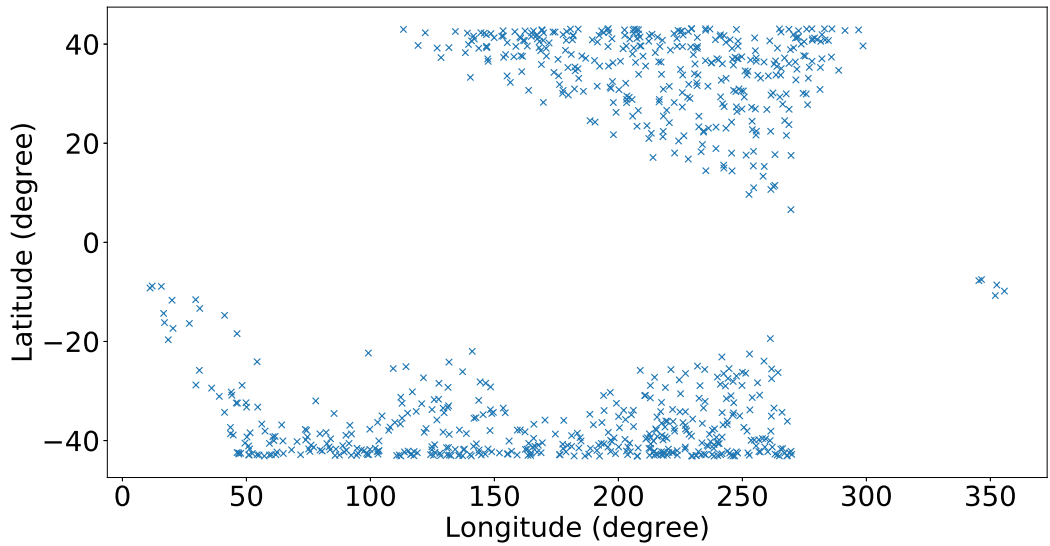}
    \caption{Geographical locations of the background flares.}
    \label{flare_map}
\end{figure}

\begin{figure}
    \centering
    \includegraphics[scale=0.4]{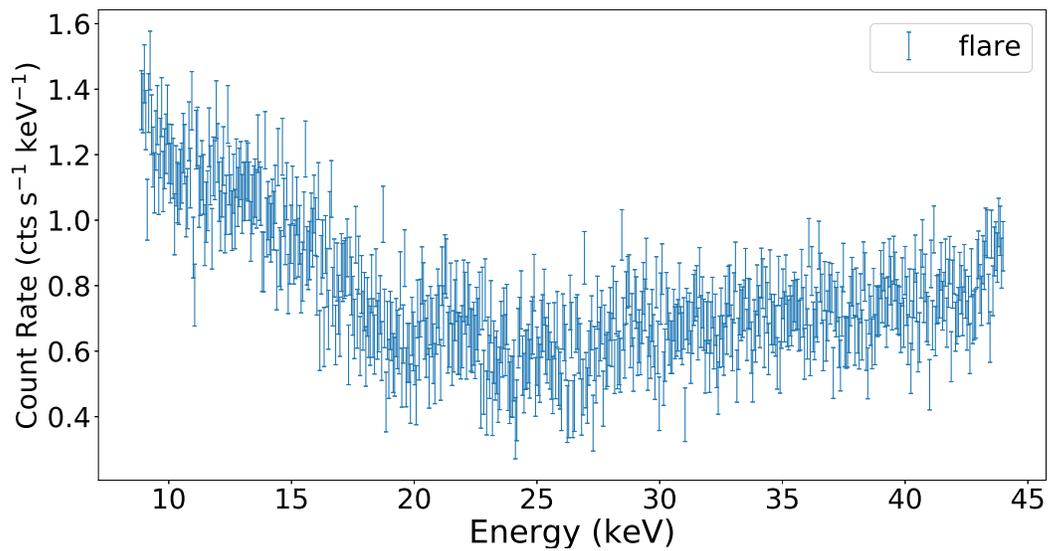}
    \caption{The average spectrum of all ME background flares. For every background flare, the spectrum is derived by subtracting spectrum of blind FOV detector from that of small FOV detector.}
    \label{flare_spectrum}
\end{figure}

After the basic GTI selection, there are still some abnormal events thus the data must be further reduced. As shown in Figure~\ref{figgrosslc}, strong flares can be seen in the light curves measured by the detectors with different FOVs. Unlike the high background rate in the low COR regions, the most obvious character of the background flare is that the flux of the flare is proportional to the FOV. 
%new1%
In addition, the flare usually lasts tens to several hundred seconds and can occur in some specific area even with the high COR (Figure~\ref{flare_distribution} and \ref{flare_map}) and can happen anytime even with the high ELV (Earth elevation angle). Therefore, the flare is suspected to be induced by the low-energy charged particles in part area of the low earth orbit, which are diffuse and can hardly penetrate the aluminum plate above the blind FOV detector and spacecraft shell. For a regular pointing observation, these flares can be identified by comparing the light curves measured by the detectors with small and large FOVs (Figure \ref{figgrosslc}), and hence be removed with the same method as adopted in LE (Liao et al. 2020). This method can detect the relatively bright flares, while the missed faint flares may slip into the filtered data and contribute the systematical error of the background model at low and high energy ends of the spectrum. Hence the difference between the spectra of the large and small FOV detectors can be attributed to the flare spectrum (Figure~\ref{flare_spectrum}). The selected flares have a mean duration $\sim200$ s and account for $\sim7\%$ of the exposures with normal GTI criteria (Table 1).
%new1%

%new1%
With two-year in-orbit operation of \emph{Insight-HXMT}, we find the noises of ME pixels vary with time. The noises sometimes become very high with the peak values up to thousands times of normal values in the range of $\rm0th-100th$ energy channel ($3.0-8.9$ keV) and take exponentially decay, which makes it easy to be detected.
%new1%
Therefore, it is necessary to perform good pixel selection. With more than 195 blank sky observations during 2017-09-20 to 2019-04-29 ($\sim1$ Ms), the typical intensity and shape of the background spectrum of each pixel can be obtained. For a blank sky observation, by comparing the detected spectrum and the typical spectrum in low energy band, the pixels temporarily suffering from the noise peak can be found and then excluded from the data analysis. There are usually $1-2$ pixels excluded in each observation. All the data of blank sky observations are reduced to build the ME background database, and then used to construct the ME background model.

\section{Characteristics of the Blank Sky Observation}
\begin{figure}
    \centering
    \includegraphics[scale=0.5]{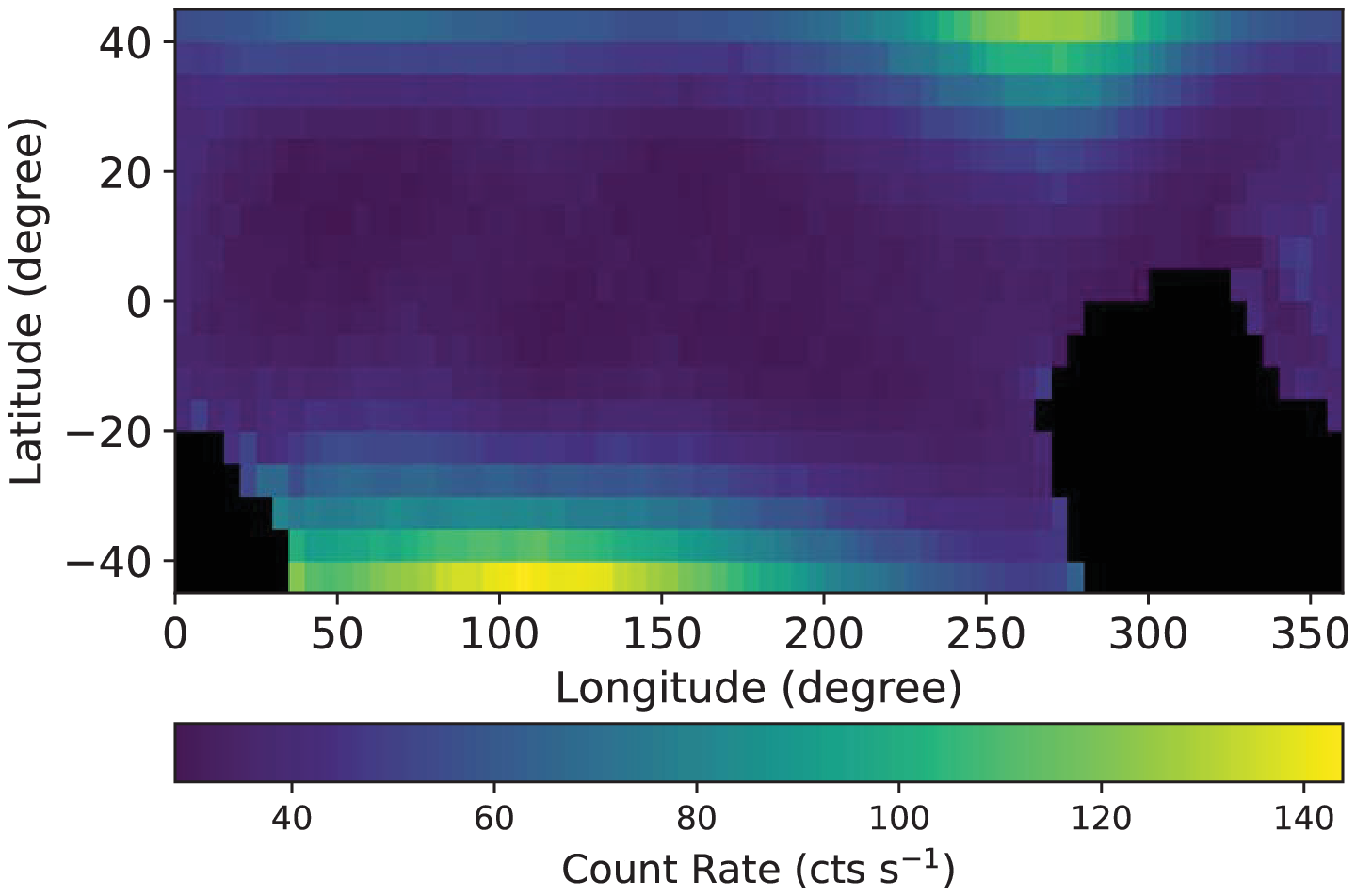}
    \includegraphics[scale=0.5]{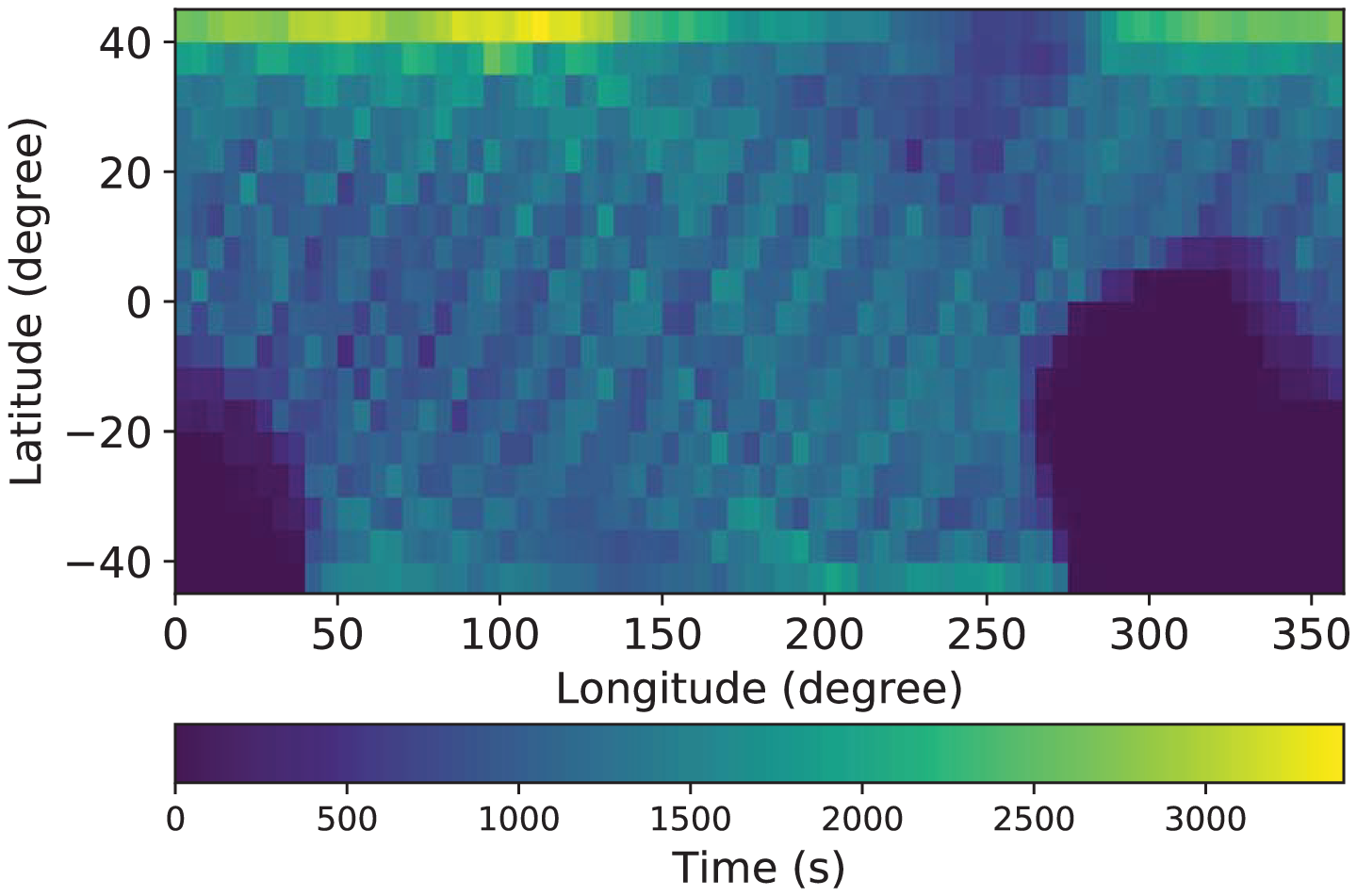}
    \caption{Geographical distribution of the background intensity of the small FOV detector in $8.9-44.0$ keV (left) and the exposure time (right).}
    \label{bkg_expt_map}
\end{figure}

%\section{Observations of the Blank Sky}
Since \emph{Insight-HXMT} has an almost circular orbit with an altitude of 550~km and a relative high inclination of $43^{\circ}$, ME passes various geographical regions characterized with different particle properties. 
Figure~\ref{bkg_expt_map} shows the geographical distribution of the ME background. The blank region is South Atlantic Anomaly (SAA), during the passage of which \emph{Insight-HXMT} is switched off to protect the instruments from the radiation damage caused by the extremely high flux of the charged particles. The background level does not increase significantly after each SAA passage, indicating that the SAA-induced background is relatively weak.
The ME background can vary by a factor of 3 with COR:$\sim30~{\rm cts}~{\rm s}^{-1}$ near the equator with the highest COR and $\sim150~{\rm cts}~{\rm s}^{-1}$ in high latitude region with the lowest COR. \emph{Insight-HXMT} passes through the equator and the high latitude region twice for each orbit of $\sim$95~minutes.

\begin{figure}
    \centering
    \includegraphics[scale=0.4]{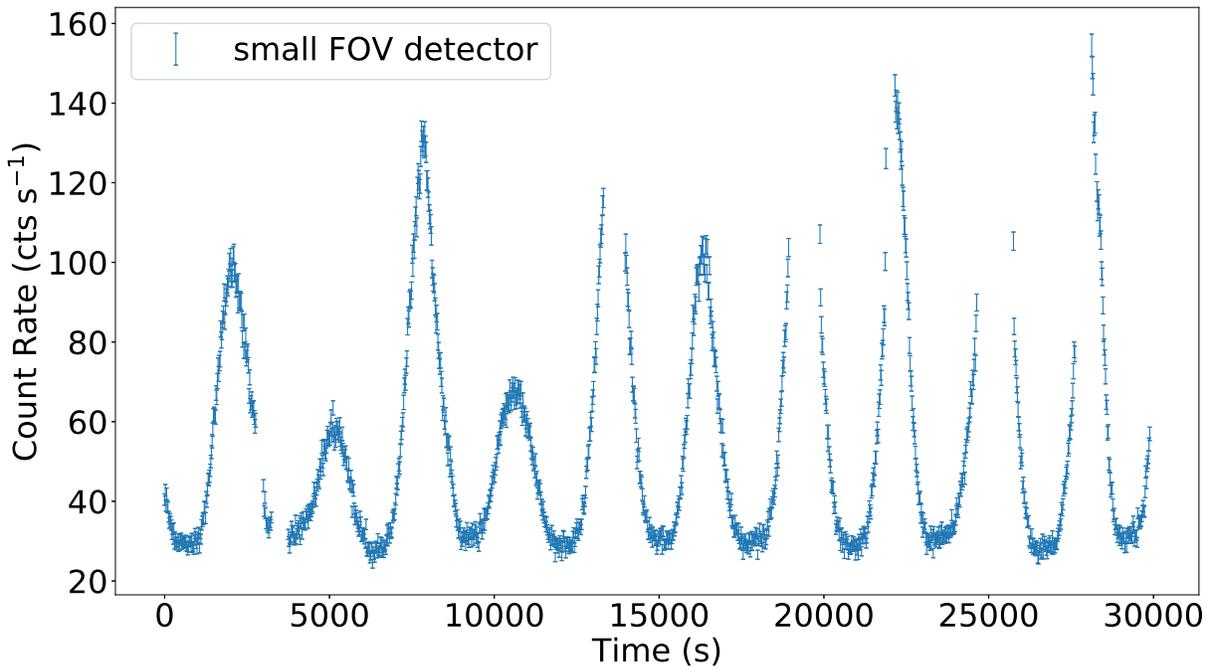}
    \caption{A typical light curve of the \emph{Insight-HXMT}/ME background in $8.9-44.0$ keV with $\rm T_{bin}=20$~s. 
    The parameters in GTI selection are described in chapter 2.1 but without the COR judgment.}
    \label{typical_lc}
\end{figure}

The light curve of a blank sky observation in the energy band $8.9-44.0$ keV is shown in Figure~\ref{typical_lc}. Such energy threshold is set to keep away from the occasional noise peaks in the lower energy band and the unstable high background in the higher energy band. According to the on-ground simulation made by Li et al. (2015), the background is mainly caused by cosmic ray protons (CRP), cosmic X-ray background through the aperture (CXB\_A) and outside the FOV (CXB\_N), albedo gamma-ray (Albedo) and SAA-induced radioactive isotopes. The CRP dominate the background and vary with COR in different geographical locations, nevertheless, CXB\_A, CXB\_N, and Albedo are independent of geographical location. 
From the on-ground simulation, the ratios of the background caused by CRP, albedo gamma-ray, CXB\_N and CXB\_A to the total background are about $60\%$, $10\%$, $15\%$ and $10\%$, respectively, in ME detection energy band, and correlate with the input environmental parameters. The SAA-induced background rises during the SAA passage and decays afterward. However, its contribution to the entire background is tiny.

\begin{figure}
    \centering
    \includegraphics[scale=0.4]{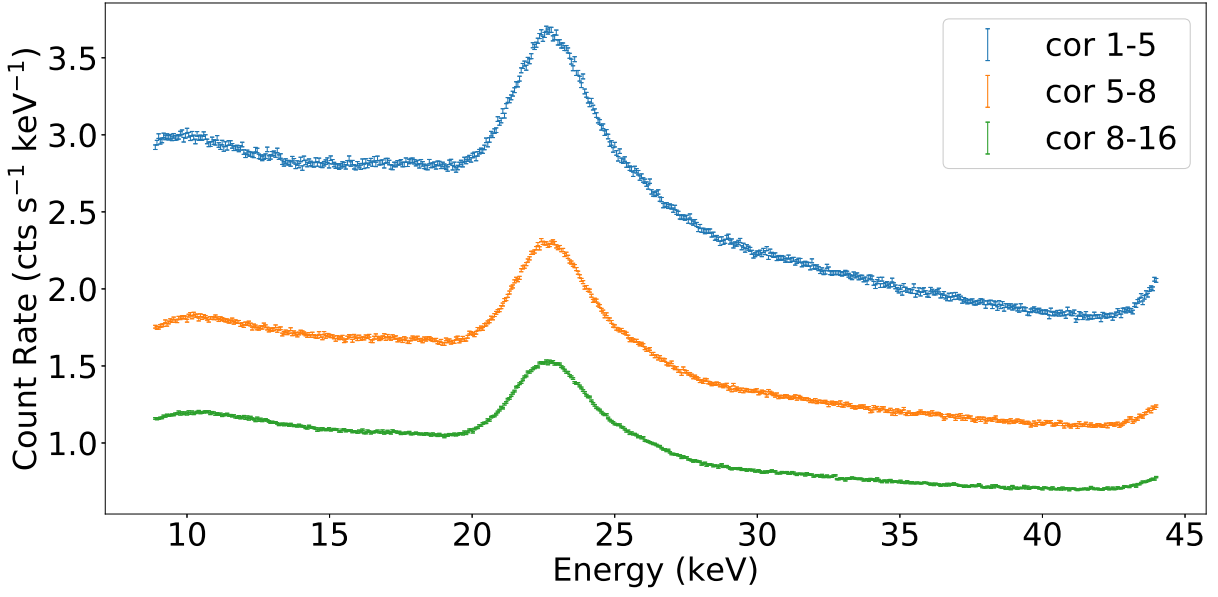}
    \caption{The ME spectra of a blank sky observation in the high Galactic latitude with the small FOV in three COR ranges.}
    \label{bkg_spec_cor}
\end{figure}

The ME spectra of the high Galactic latitude blank sky observations measured with the small FOVs in three COR regions are shown in Figure~\ref{bkg_spec_cor}. There are no obvious emission lines except the silver fluorescence line at $\sim22$~keV, which originates from the silver glue under Si-PIN pixels. Although the background intensity strongly depends on geographical location, the spectral shape is rather stable in the same geographical location (Section 5 for details). The spectra in the high COR regions tend to have larger slopes and relatively stronger silver fluorescence line. %Since the Ag fluorescence line is mainly stimulated by the CXB from aperture or through the shell of spacecraft, it is less influenced by the local particle environment.

\section{Method and Test of the Background Modeling}

\subsection{Model Method}
The background of \emph{Insight-HXMT}/ME mainly has the following features: 
both the intensity and spectra shape vary with the geographical location; 
in the same geographical location, the spectral shape is stable and the intensity can vary with the flux of the charged particles; 
there is no long-term evolution in ME background.
Since the background of \emph{Insight-HXMT}/ME is correlated with geographical location, a background database can be built as the essential input of background modelling. The entire geographical region covered by 
\emph{Insight-HXMT} ($0^{\circ}<{\rm long.}<360^{\circ}$; $-43^{\circ}<{\rm lat.}<43^{\circ}$) is divided into hundreds of small grids ($5^{\circ}\times5^{\circ}$ 
in geographical coordinate system). In each grid, the local space environment is stable and shows little fluctuation (Tawa et al. 2008). 
With the two-year blank sky observations of \emph{Insight-HXMT}, the typical spectra (spectral shape and intensity) of both the small and blind FOV detectors in each grid can be obtained. 
For a pointing observation, the satellite passes through \emph{n} grids, the estimated background flux (as a function of energy channel $c$) in the $i$-th grid $F_{\rm est}(c;i)$ can be calculated as
\begin{equation}
F_{\rm est}(c;i) = f_{i} F_{\rm exp}(c;i), \\
f_{i}=\frac{C_{\rm obs}({i|\rm BD})}{C_{\rm exp}({i|\rm BD})},
\end{equation}
where $f_{i}$ is the correction factor calculated as the ratio of the observed and the expected count rates of the blind FOV detector in the energy range of $8.9-44.0$ keV, and BD means the blind FOV detector. The estimated background $F_{\rm exp}(c)$ can be calculated as
\begin{equation}
F_{\rm exp}(c)=\frac{\sum_{n}F_{\rm est}(c;i)T_{i}}{\sum_{n}T_{i}},
\end{equation}
where $T_{i}$ is the effective exposure in the $i$-th grid.

\subsection{Model Test}
The blank sky observations are also used to test the background model. In order to keep the independence between data and model, the blank sky observations are divided into two parts and each has an exposure of ${\rm T_{exp}}=500$~ks which amounts to half of the total background observational time. The first half are used to build the background database as input to background modelling, while the second half are used to test the derived background model. Three types of model test are carried out and shown as follows. 

\begin{figure}
    \centering
    \includegraphics[scale=0.4]{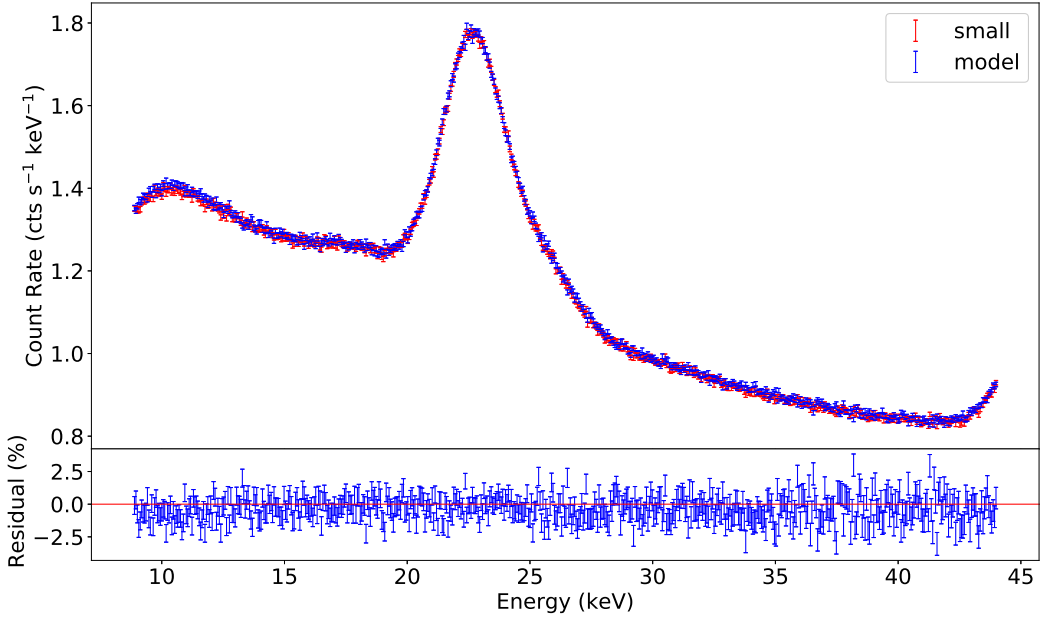}
    \caption{Comparison of the observed and estimated background spectrum with the exposure time $\rm T_{exp}=500$~ks.}
    \label{comparison_500ks}
\end{figure}

The first test is to take extremely long exposure time. The observed spectrum is obtained by merging all the $500$~ks blank sky observations, and the estimated background spectrum is obtained with the background model. 
Figure~\ref{comparison_500ks} shows the comparison of the observed and estimated background spectra under an exposure $\rm T_{exp}=500$~ks. The red and blue data, overlapped in the top panel of Figure~\ref{comparison_500ks}, represent the observed spectrum and the estimated background, respectively. The residuals are shown in the bottom panel and there is no obvious structure in the residuals except, indicating the success of the background model adopted. The average ratio of the residual to the background is $0.153\pm0.015\%$ in the whole energy band ($8.9-40.0$~keV).
Because of various uncertainties (e.g., low-significance flare, electronic noise), there are systematic errors and biases in background estimation. Figure~\ref{comparison_500ks} shows one example of the high-statistical test of background model, where the weak biases are visible in several energy bands. Both the flares and the variances of the cosmic-ray protons can be the causes of the biases. Although the bias can in principle be corrected, since the biases are much smaller than the systematic error, the influence is ignored in the current background modelling.

\begin{figure}
    \centering
    \includegraphics[scale=0.4]{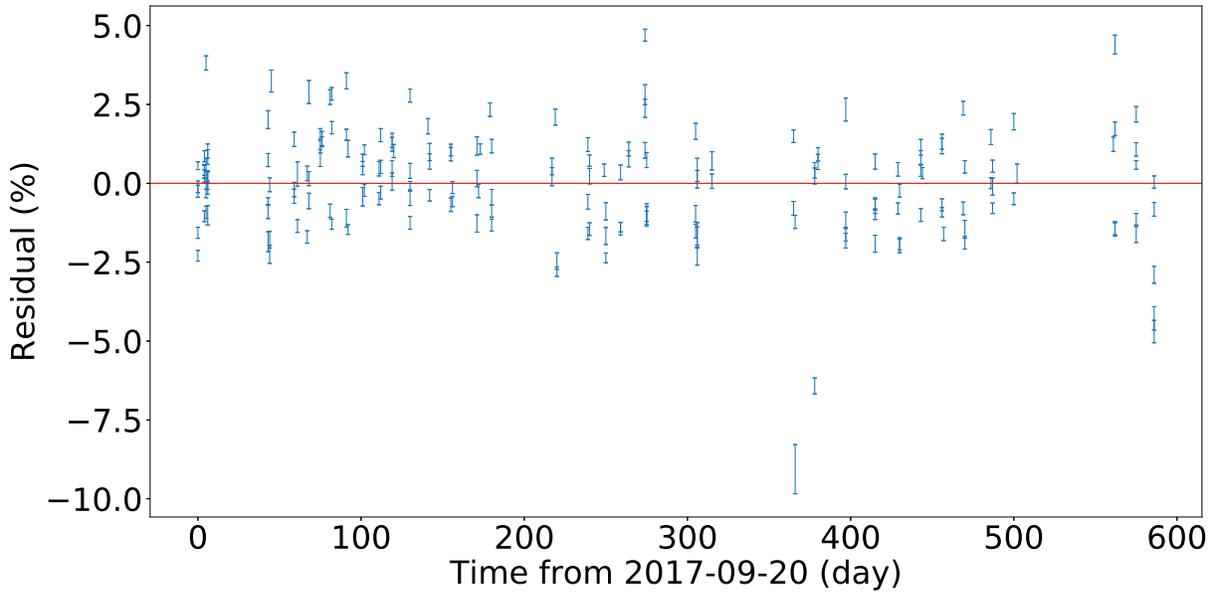}
    \caption{Residuals of the observed and estimated count rates of 195 blank sky observations in $8.9-44.0$ keV.}
    \label{res_every_bls}
\end{figure}

\begin{figure}
    \centering
    \includegraphics[scale=0.4]{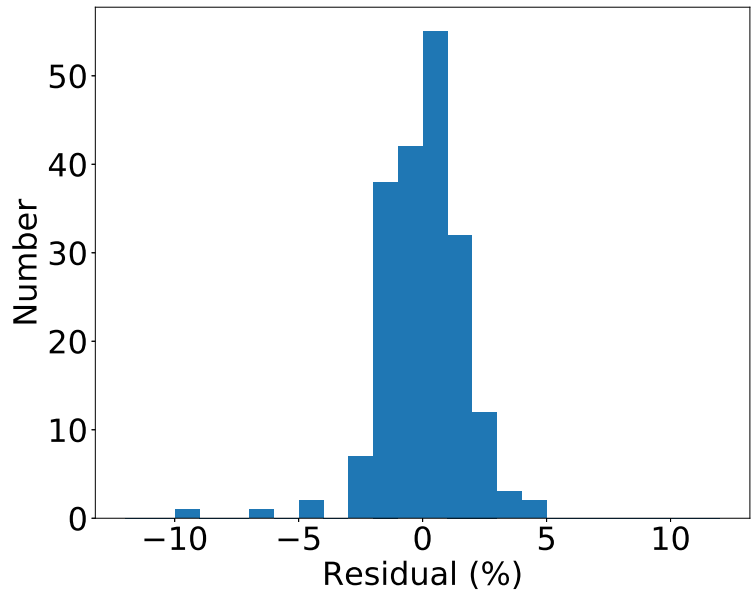}
    \caption{Distribution of the residuals shown in Figure~\ref{res_every_bls}}
    \label{dis_res}
\end{figure}

\begin{figure}
    \centering
    \includegraphics[scale=0.4]{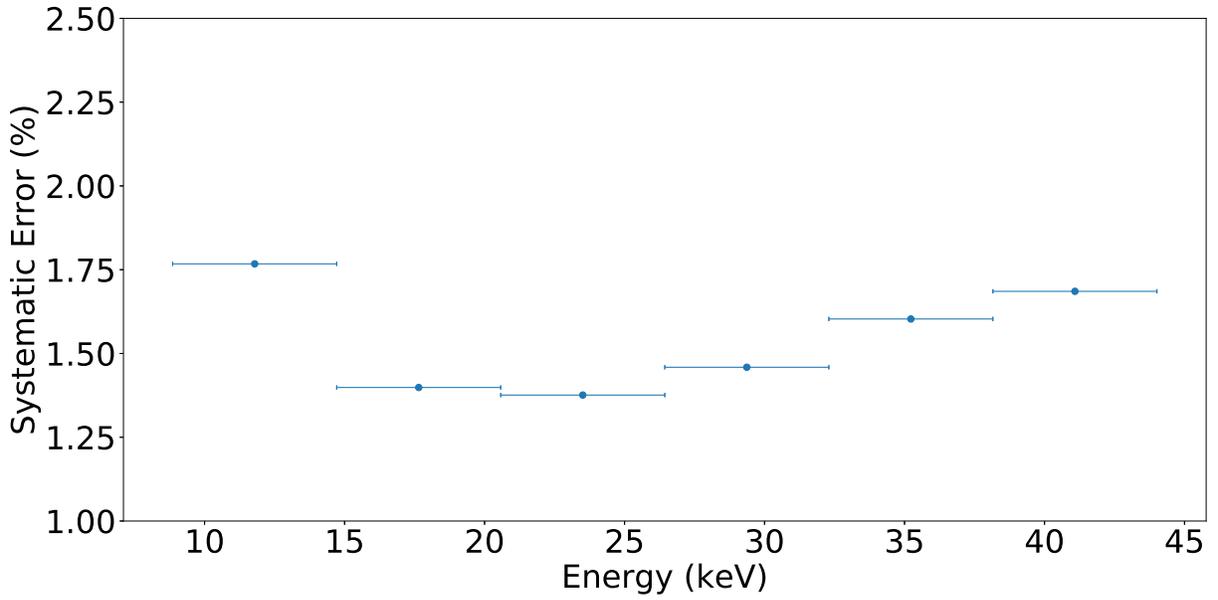}
    \caption{Systematic errors of background model in six energy bands that shown in Table 2.}
    \label{sys_err_energy}
\end{figure}

The second test is to take each blank sky observation enclosed in the test data.
For each black sky observation, the background spectrum can be modelled and compared to the observed one.
With the method in Fukazawa et al. (2009), the systematic error $\sigma_{\rm sys}$ can be calculated by
\begin{equation}
\sigma_{\rm sys} = \sqrt{\sum_{i} \frac{1}{\omega_{i}} (\sigma_{{\rm t},i}^{2}-\sigma_{{\rm stat},i}^{2})}, 
\end{equation}
where
\begin{equation}
\omega_{i} = \frac{1}{\sigma_{{\rm stat},i}^{2}} \Sigma_{i}\left(\frac{1}{\sigma_{{\rm stat},i}^{2}}\right),
\end{equation}
where $\sigma_{{\rm t}, i}$ and $\sigma_{{\rm stat}, i}$ are the total dispersion and the statistical errors of the residuals, 
and $\omega_i$ refers to the weight factor of the $i$-th observation.
Figures~\ref{res_every_bls} and \ref{dis_res} show the residuals of each observation and the residual distribution.
The exposure time averaged over each blank sky observation in this test is $\sim5.5$~ks, and the resulted systematic error is 1.3\% in the full energy band.
The relationship between the systematic error and the energy band is also investigated.
Some specious biases can also be caused by the weak flares and  the variances of the cosmic-ray protons.
As shown in Figure~\ref{sys_err_energy}, the energy band is divided into six intervals uniformly, and the systematic errors are in general $<2\%$. 
The systematic error in the lowest energy band is relatively larger, probably due to the unfiltered noise peak and the worse energy resolution in the lower energy band. 
It is worth noting that the relationship between the systematic errors and energy is somewhat similar to the spectral shape of the flares (Figure~\ref{flare_spectrum}).
This may suggest that the systematic error can be partly due to the flares with low statistical significance.

\begin{figure}
    \centering
    \includegraphics[scale=0.4]{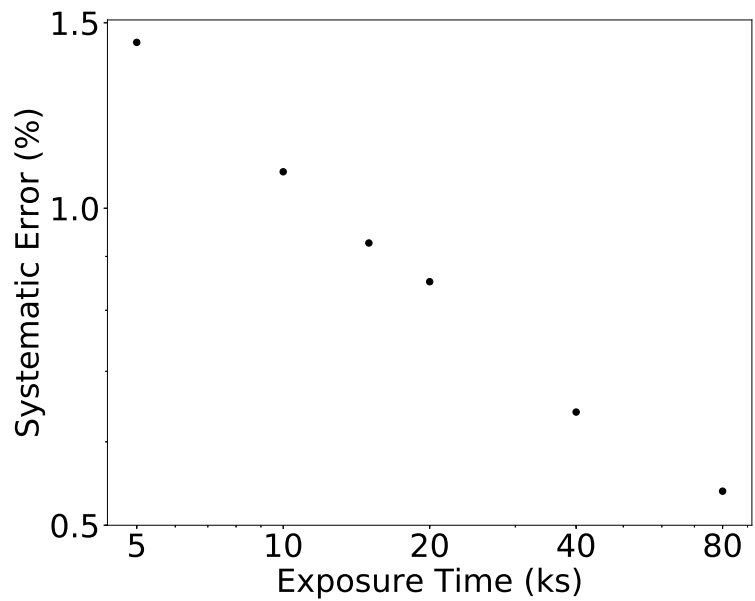}
    \caption{Relationship of the systematic errors of background model and the exposure times in $8.9-44.0$ keV.}
    \label{sys_err_expt}
\end{figure}

The third test is to investigate the systematic error with different exposure times.
All the test data are merged together and then re-grouped into several sub-observations with the same exposure time.
For each sub-observation an analysis similar to the second test is performed to estimate the systematic error.
Accordingly, the systematic errors under exposures of ${\rm T_{exp}}=5,10,15,20,40$ and 80~ks are obtained and, as shown in Figure~\ref{sys_err_expt}, there is an obvious anti-correlation 
between the systematic error and the exposure time.

\section{Discussion}
\begin{figure}
    \centering
    \includegraphics[scale=0.4]{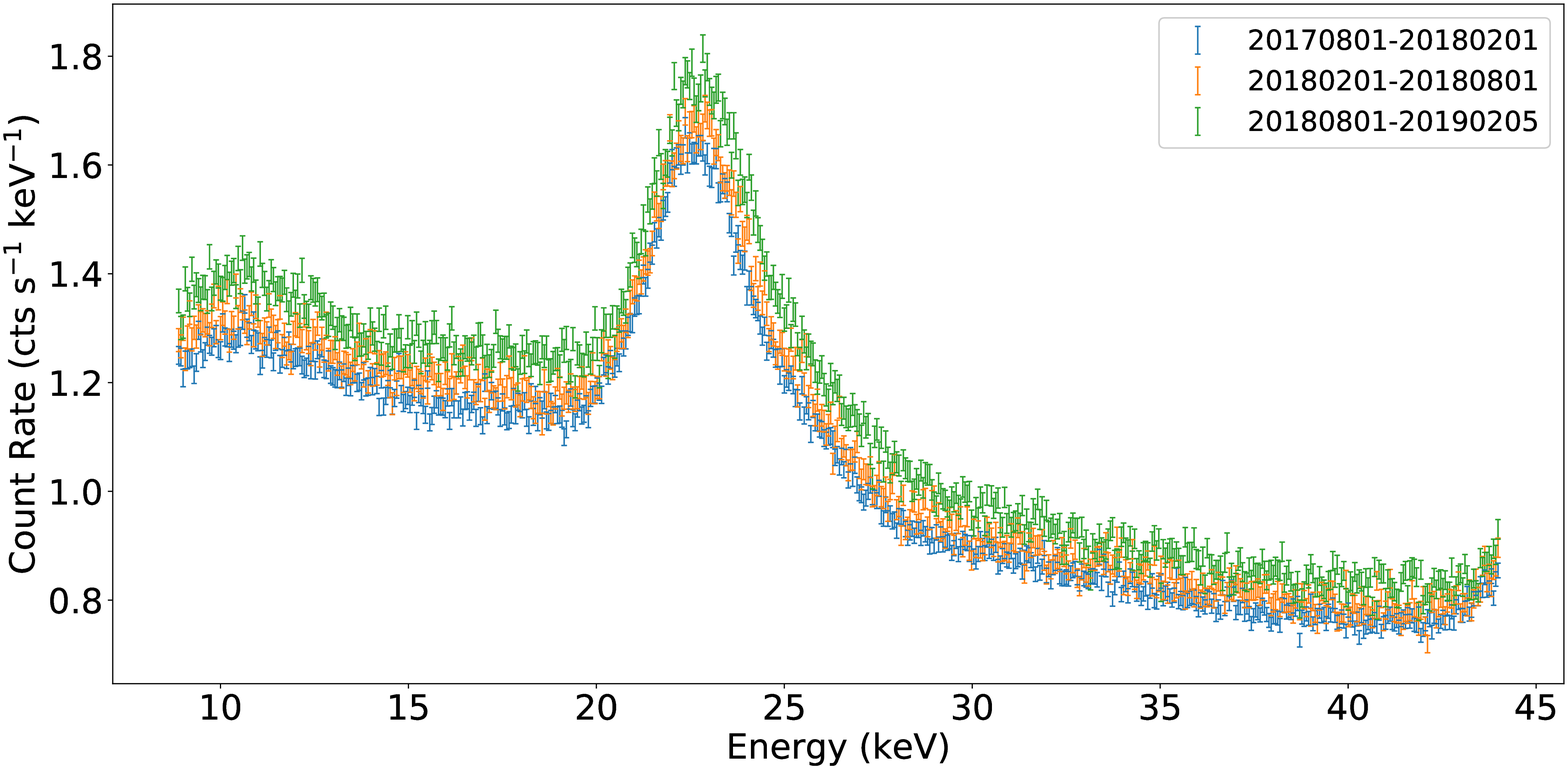}
    \includegraphics[scale=0.4]{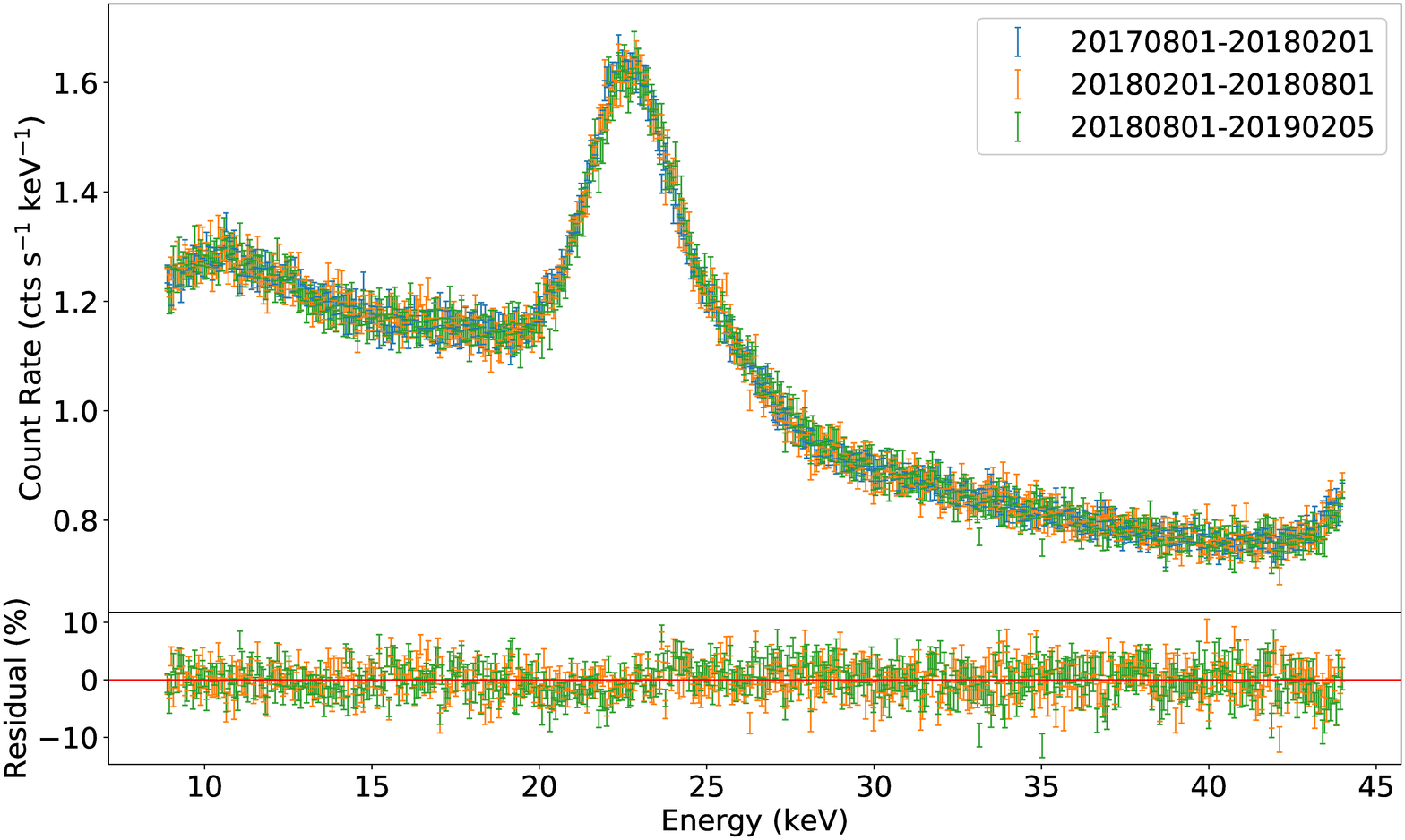}
    \caption{Comparison of spectra of the blank sky observations in different epochs. The COR range is form $8-10$ GeV.
    The bottom panel shows the normalized and the difference between these in the first and the other two epochs.}
    \label{spec_time}
\end{figure}

\begin{figure*}
\gridline{\fig{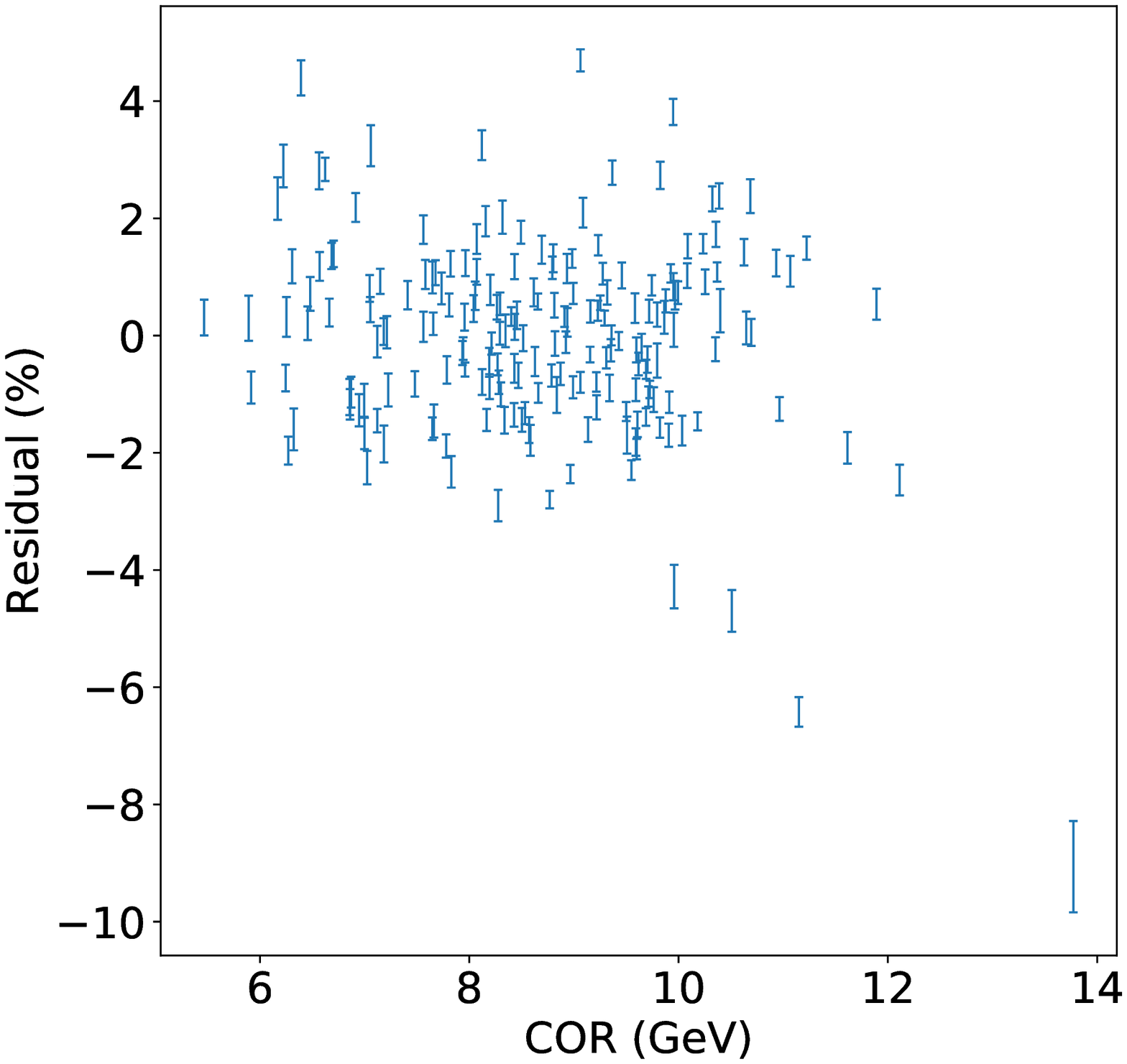}{0.3\textwidth}{(a)}
          \fig{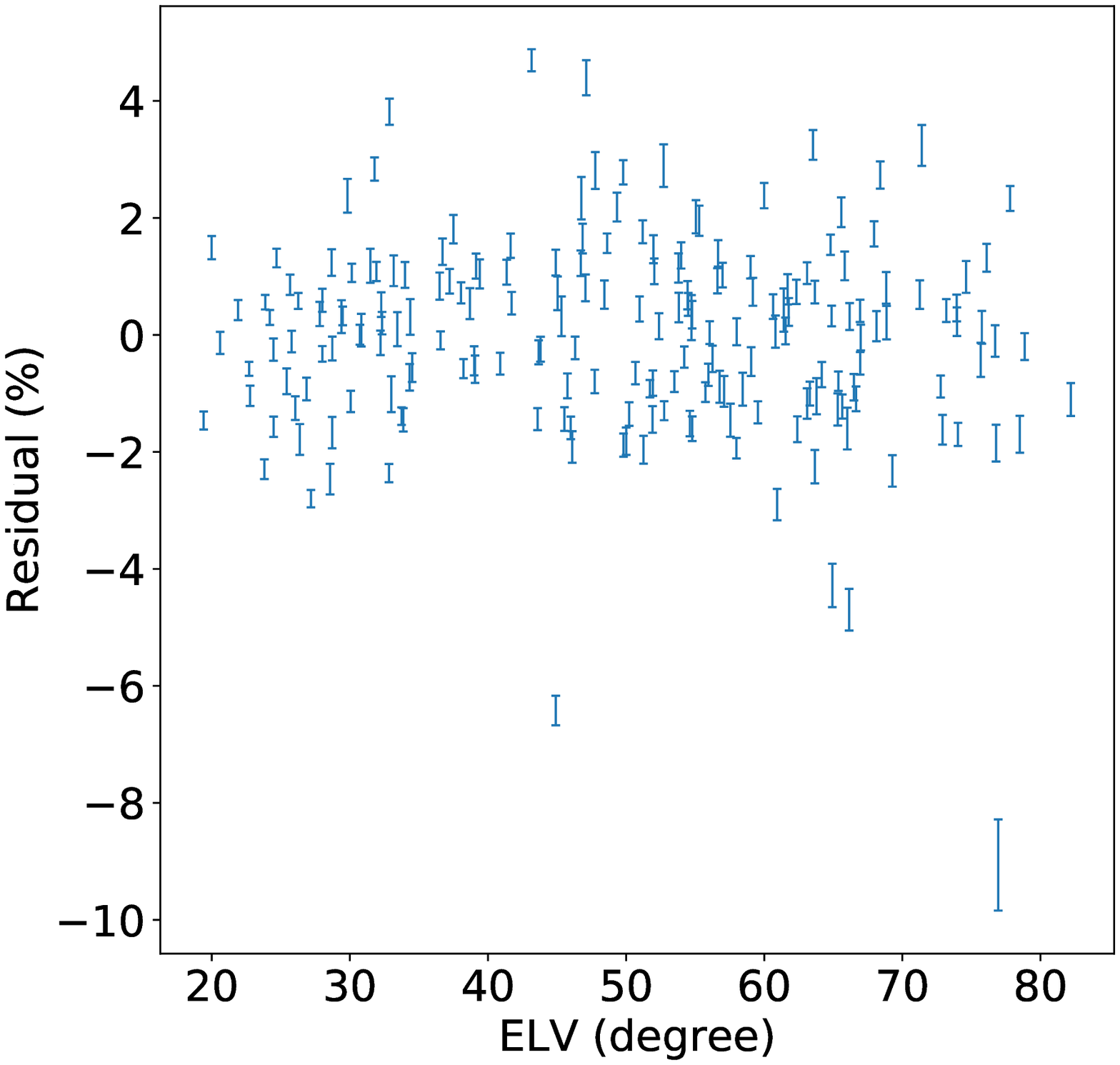}{0.3\textwidth}{(b)}
          \fig{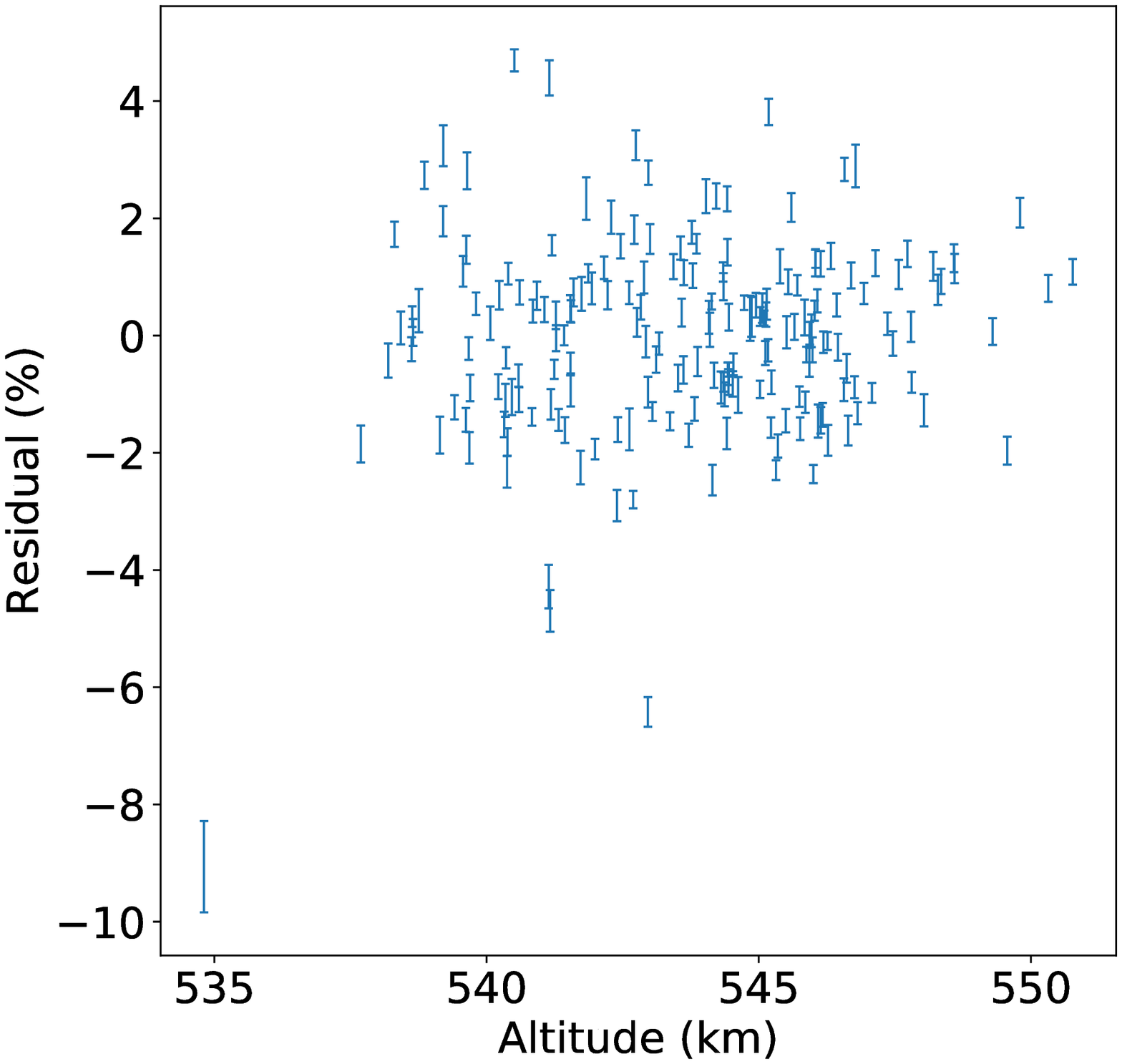}{0.3\textwidth}{(c)}
          }
\gridline{
          \fig{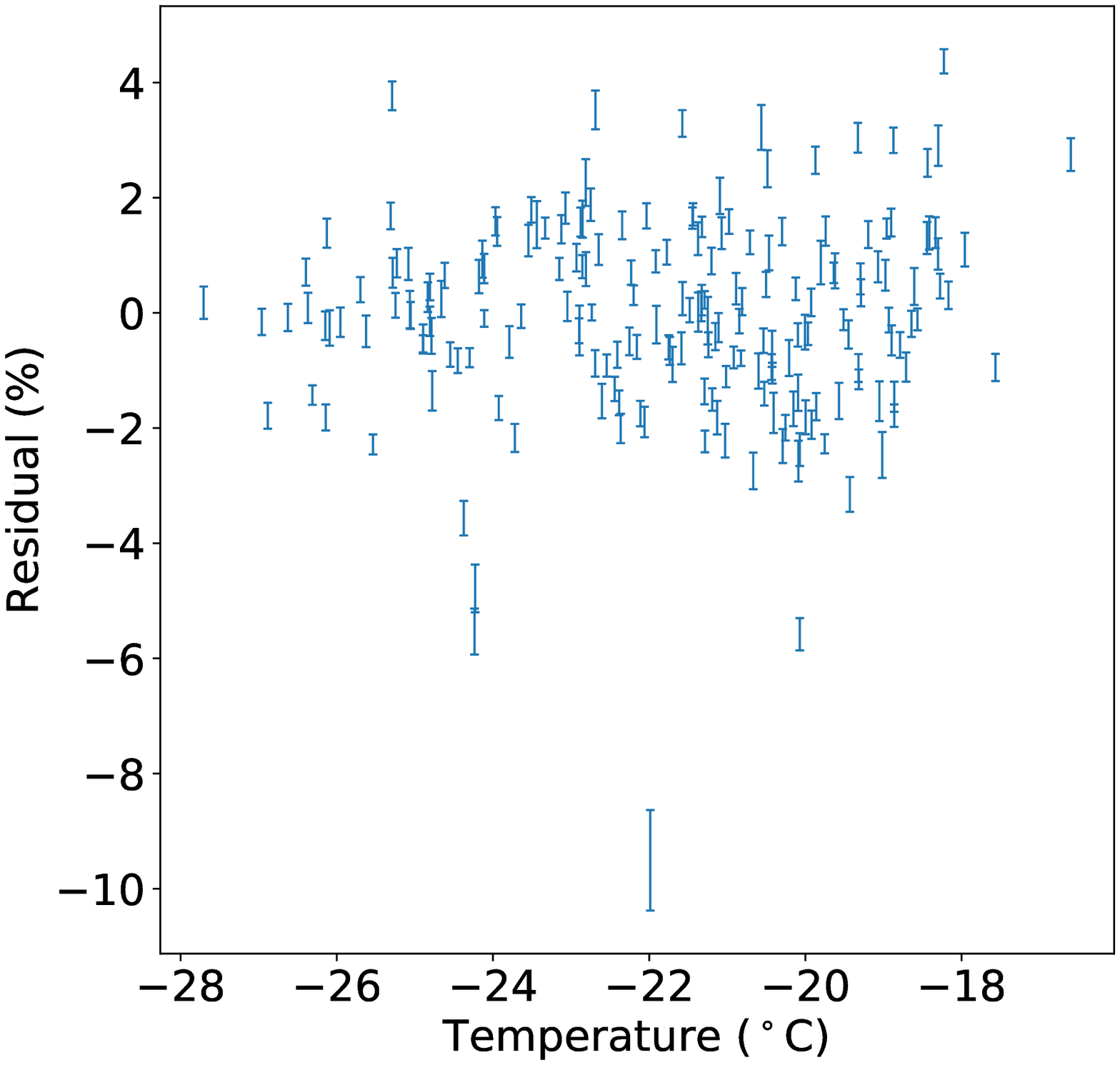}{0.3\textwidth}{(d)}
          \fig{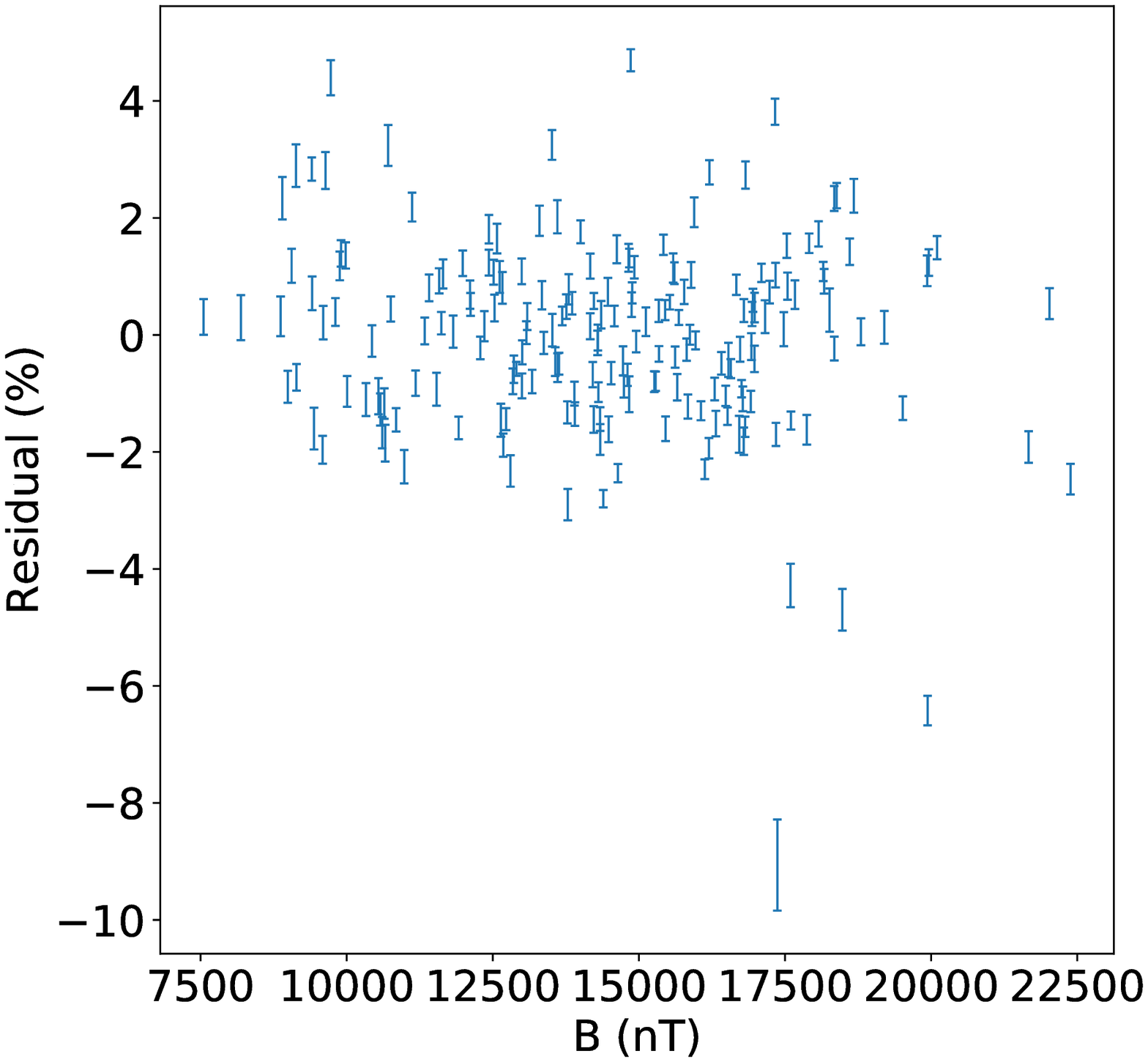}{0.3\textwidth}{(e)}
          }
\caption{Relationship of the residuals of the model test in $8.9-44.0$ keV and the COR (a), ELV (b), satellite altitude (c), instrumental temperature (d) and the local magnitude of geomagnetic field (e). The values of each point in each panel are the average values of the parameters in a blank sky observation.}
\label{res_cor_elv_alt}
\end{figure*}

%\subsection{spectra in the same geographical location}
In ME background modelling, a critical assumption is that the spectral shape is stable for each grid. 
However, due to the relatively small grid size and low ME background count rate, the data are insufficient to test this assumption. Therefore, we combine the data with similar COR in order to enlarge the data sample. The data of those grids with COR value varying within $8-10~{\rm GeV}$ 
are divided into three parts according to the observational time and then the spectra in three epochs are obtained to make the comparison. As shown in Figure~\ref{spec_time}, the intensities of the three spectra are different but the spectral shapes are almost consistent with each other.
%\subsection{Exposure time}
%In the background model, the blind FOV detector is used to correct the background intensity in each grid, 
The blind FOV detector is used to correct the background intensity in each grid, 
therefore, the statistical error of the blind FOV detector will be transformed into the systematic error of the estimated background. As the exposure time increases, the statistic error of the blind FOV detector decreases, 
and the systematic error of the estimated background decreases accordingly (Figure~\ref{sys_err_expt}).
%\subsection{Parameter Dependence}
Currently, only the geographical location and the count rate of the blind FOV detector are considered in the background estimation. Figure~\ref{res_cor_elv_alt} shows the relationship of the residuals with COR, ELV, 
satellite altitude, instrumental temperature and the local magnitude of geomagnetic field. There are no obvious correlations in all panels of Figure~\ref{res_cor_elv_alt}, 
which means that it is reasonable to ignore the influence from the COR, ELV, satellite altitude in the current background modelling.

%\subsection{Galactic ridge X-ray emission}
In this paper, the background modelling is based on the background database, which is built with more than 195 observations of the high Galactic latitude. This means that the estimated background is mainly composed of the particle component and the CXB. 

Revnivtsev et al. (2003) reported the CXB has a fractional fluctuation $7\%$ per square deg. With the CXB spectral parameters, the ME instrumental response and the observed blank sky spectra, the CXB roughly account for $10\%$ of the total ME background. The FOVs of the small and large FOV detectors are 4 and 16 square degrees, respectively, thus the fluctuation of the CXB contribute uncertainty of $\sim 0.4\%$ and $\sim 0.2\%$ in ME background estimation in 10-30 MeV band. These fluctuations are much smaller than the systematic error and thus the influence of the CXB fluctuation can be ignored in the current background modelling.

%new1%
However, also due to relatively large FOV, the Galactic ridge X-ray emission (GRXE) will inevitably be a part of ME background for the observation of the Galactic regions. A simulation with the spectral GRXE parameters from Galactic bulge ($-30^{\circ}<l<30^{\circ}~\&~-15^{\circ}<b<15^{\circ}$) 
given by INTEGRAL (Türler et al. 2010) and the response of \emph{Insight-HXMT}/ME is performed to investigate the impact of the GRXE on the background.
We find that the GRXE will mainly affect the background with the energy lower than 26~keV and becomes more serious at 11.1~keV.
For the whole energy range of $8.9-44.0$ keV, the GRXE contributes less than 3\% of the total background, 
which is slightly larger than the typical systematic error derived with the current background model.
This indicates that for the source in the Galactic center, the systematic uncertainty is mainly composed of two parts: 
one is the systematic error of the particle background and another is the GRXE.
Since the intensity of GRXE is greater than the systematic error derived based purely on the particle background, 
it is possible to measure GRXE with the Galactic plane survey of \emph{Insight-HXMT}/ME, which is beyond the scope of this paper and will be reported elsewhere.

\section{Summary}
In this work, a database of \emph{Insight-HXMT}/ME has been built with the two-year blank sky observations. 
Based on the database, we had established the background model of \emph{Insight-HXMT}/ME. The spectral shape of the ME background is obtained by averaging over all the geographical grids experienced by the satellite during an observation, weighted by the contemporary count rate measured by the blind FOV detector. The blank sky observations are also used to test the model and estimate the systematic error. 
The current reproducibility of the ME background modelling is estimated to be better than $1.3\%$ ($8.9-44.0$~keV) 
for an individual observation with a typical exposure time of 5.5~ks. 
In addition, the systematic error is anti-correlated to the exposure time, e.g., the systematic error can be $<0.5\%$ for an observation with the exposure time $\rm T_{exp}>80$~ks. Since the statistical error in measurement of the count rate by the blind FOV detector can contribute partially to the systematic error of the estimated background, the entire systematic error of the current background model can be as larger as $~3\%$ for an observation with $\rm T_{exp}<1$~ks. 
Since the current background modelling highly relies on the blind FOV detector, a parameterized background model, which is free from the blind FOV detector, is being developed in parallel in a way similar to that adopted in RXTE (Jahoda et al. 2006).

\acknowledgments{This work made use of the data from the \emph{Insight-HXMT} mission, a project funded by China National Space Administration (CNSA) and the Chinese Academy of Sciences (CAS). The authors thank supports from the National Program on Key Research and Development Project (Grant No. 2016YFA0400801, 2016YFA0400802) and the National Natural Science Foundation of China under Grants No. U1838202 and U1838201.}

\bibliographystyle{plainnat}

\begin{table}[ptbptbptb]
\begin{center}
\label{table0}
\caption{Parameters in GTI selection.}
 %\vspace{5pt}
\begin{tabular}{cchlDlc}%{cccccccccccccccccc}
\\\hline
 Parameters     & Criteria \\\hline
 $\rm ELV$         &    $\rm >5^{\circ}$      \\       
 $\rm DYE\_ELV$  &    $\rm >10^{\circ}$   \\         
 $\rm COR$        &    $\rm >5^{\circ}$    \\
 $\rm SAA\_FLAG$   &   $\rm =NO$          \\
 $\rm T\_SAA$  &       $\rm >100\ s$      \\    
 $\rm TN\_SAA$   &      $\rm >100\ s$     \\
 $\rm SUN\_ANGLE$   &    $\rm >30^{\circ}$      \\
 $\rm MOON\_ANGLE$   &   $\rm >30^{\circ}$       \\
 $\rm ANG\_DIST$   &     $\rm >0^{\circ}.02$   \\
\hline
\end{tabular}
\end{center}
\begin{list}{}{}
\item[NOTE:]{ELV: earth elevation of the FOV center direction. $\rm DYE\_ELV$: day earth elevation of the FOV center direction. COR: geomagnetic cut-off rigidity in unit of GeV. $\rm SAA\_FLAG$=NO: Not in SAA. $\rm T\_SAA$: Time (s) after SAA (south Atlantic anomaly) passage. $\rm TN\_SAA$: Time (s) to next SAA passage. $\rm ANG\_DIST$: Difference of the real pointing direction from the target position. $\rm ANG\_DIST <0^{\circ}.02$ means stable pointing.}
\end{list}
\end{table}

\begin{table}[ptbptbptb]
\begin{center}
\label{table1}
\caption{The systematic errors of the background estimation in six energy bands}
 %\vspace{5pt}
\begin{tabular}{cchlDlc}%{cccccccccccccccccc}
\\\hline
 Energy Band (keV)     & Systematic Error \\\hline
 $\rm 8.9-14.7$         &    $\rm 1.8\%$      \\       
 $\rm 14.7-20.6$  &    $\rm 1.4\%$   \\         
 $\rm 20.6-26.4$        &    $\rm 1.4\%$    \\
 $\rm 26.4-32.3$   &   $\rm 1.5\%$          \\
 $\rm 32.3-38.2$  &       $\rm 1.6\%$      \\    
 $\rm 38.2-44.0$   &      $\rm 1.7\%$     \\
\hline
\end{tabular}
\end{center}
\end{table}

%new1%

%new1%

\end{document}